\documentstyle{article}

\begin{document}

   \title{Scintillation in scalar-tensor theories of gravity}
\author{ C. Bracco\\
Observatoire de Haute-Provence,CNRS, F-04870 
Saint-Michel l'Observatoire, France.\\
P. Teyssandier\\
 Laboratoire de Gravitation et Cosmologie Relativistes, 
Universit\'e Pierre et Marie Curie\\
CNRS/ESA 7065, Tour 22/12, BP 142, 4 place Jussieu, F-75252 Paris Cedex 05, 
France.}
 
   \date{Received 30 July 1998/Accepted 09 september 1998}

   \maketitle

   \begin{abstract}
 
We study the scintillation produced by time-varying gravitational fields within
scalar-tensor theories of gravity. The problem is treated in the geometrical
optics approximation for a very distant light source emitting quasi plane 
monochromatic
electromagnetic waves. We obtain a general formula giving the time dependence
of the photon flux received by a freely falling observer. In the weak-field
approximation, we show that the contribution to the scintillation effect due to
the focusing of the light beam by a gravitational wave is of first order in the
amplitude of the scalar perturbation. Thus
scalar-tensor theories contrast with general relativity, which predicts that
the only first-order effect is due to the spectral shift. Moreover, we find
that the scintillation effects caused by the scalar field have a local
character: they depend only on the value of the perturbation at the observer. 
This effect provides in principle a mean to detect the presence of a long range
scalar field in the Universe, but its smallness constitutes a tremendous
challenge for detection.
     

   \end{abstract}

\section{Introduction}
The plane optical wavefronts of a distant background light source become rippled
when they cross a perturbation. For a distant perturbation,
the focusing of light at the observer changes with the curvature of
the ripples. It is the usual geometrical
scintillation effect. It accounts, {\it e.g.}, for the
twinkling of stars under the atmospheric turbulence.
\par The question whether gravitational waves can cause the light emitted by a
distant source to scintillate is an old problem. 
In general relativity, it is well known from the early works by
Zipoy (1966), Zipoy \& Bertotti (1968) and Bertotti \& Trevese (1972)
that gravitational waves have no
focusing property to the first order in their amplitude. 
\par However, it has been recently pointed out by Faraoni (1996) that a
first-order scintillation effect can be expected in scalar-tensor theories of 
gravity \footnote{ On these theories, see, {\it e.g.}, Will (1993) and 
Damour \& Esposito-Far\`ese (1992), and references therein.}. 
Furthermore, some actual improvements of the observational
techniques renew the interest in the search of gravitational scintillation
(Labeyrie 1993) and related effects (Fakir 1995). 
\par The aim of the present work is to make a detailed analysis of the
scintillation effect in monoscalar-tensor theories for a monochromatic
electromagnetic wave propagating in a weak gravitational field. 
We adopt the point of view that the physical metric is the metrical tensor 
$g_{\mu \nu}$ to which matter is universally coupled. This basic assumption 
defines the usual "Jordan-Fierz" frame. We find a scintillation 
effect proportional to the value of the scalar field perturbation at the 
observer. 
\par Our result contrasts with the zero effect found by Faraoni \& Gunzig
(1998) by using the "Einstein" conformal frame, in which the original physical
metric $g_{\mu \nu}$ is replaced by a conformal one \footnote{A clear
distinction between the "Einstein" frame and the "Jordan" frame, may be found,
{\it e.g.}, in Damour \& Nordverdt (1993)}.
However, their negative conclusion is seemingly due to the fact that the
authors do not take into account the changes in areas and other physical
variables induced by the conformal transformation (Damour \&
Esposito-Far\`ese 1998). 
\par The paper is organized as follows. In Sect.2, we give the notations and we
recall the fundamental definitions. In Sect.3, we construct the theory of
gravitational scintillation for a very distant light source emitting quasi
plane electromagnetic waves. Our calculations are valid for any metric theory
of gravity in the limit of the geometrical optics approximation. We obtain the
variation with respect to time of the photon flux received by a freely falling
observer as a sum of two contributions: a change in the scalar amplitude of the
electromagnetic waves, that we call a geometrical scintillation, and a change
in the spectral shift. We express each of these contributions in the form of an
integral over the light ray arriving to the observer. In Sect.4, we study the
scintillation within the linearized weak-field approximation. We show that the
geometrical scintillation is related to the Ricci tensor only. Thus we recover
as a parti\-cular case the conclusions previously drawn by Zipoy and Zipoy \&
Bertotti for gravitational waves in general relativity. Moreover, we show that
the contribution due to the change in the spectral shift is entirely determined
by the curvature tensor. In Sect.5, we apply the results of Sect.4 to the
scalar-tensor theories of gravity. We prove that these theories predict a
scintillation effect of the first order, proportional to the amplitude of the
scalar perturbation. Furthermore, we find that this effect has a local
character: it depends only on the value of the scalar field at the observer. 
Finally, we briefly examine the possibility of observational tests in Sect.6. 

\section {Notations and definitions}

The signature of the metric tensor $g_{\mu \nu}(x)$ is assumed to be
{\scriptsize $(+~-~-~-)$}. Indices are lowered with $g_{\mu \nu}$ and raised 
with $g^{\mu \nu}$.
\par Greek letters run from 0 to 3. Latin letters are 
used for spatial coordinates only: they run from 1 to 3. 
A comma (,) denotes an ordinary partial 
differentiation. A semi-colon (;) denotes a covariant partial differentiation 
with respect to the metric; so $g_{\mu \nu; \rho} = 0$. 
Note that for any function $F(x)$, $F_{;\alpha}~=~F_{, \alpha}$.
\par Any vector field $w^{\rho}$ satisfies the following identity 
\begin{equation}
w^{\rho}_{\verb+ +;\mu ;\nu} - w^{\rho}_{\verb+ +;\nu ;\mu} = -
R^{\rho}_{. \sigma \mu \nu} w^{\sigma}  
\end{equation}
where $R^{\rho}_{. \sigma \mu \nu}$ is the Riemann curvature
tensor (note that this identity may be regarded as defining the curvature
tensor). The Ricci tensor is defined by
\begin{equation}
R_{\mu \nu} = R^{\lambda}_{. \mu \lambda \nu}
\end{equation}
\par Given a quantity $P$, $\overline{P}$ denotes its complex conjugate. 
\par The subscripts $em$ and $obs$ in formulae stand respectively for emitter 
and observer. 
\par The constant $c$ is the speed of light and $\hbar$ is the Planck constant 
divided by $2 \pi$.

\section{General theory of the gravitational scintillation}
In a region of spacetime free of electric charge, the propagation equations for
the electromagnetic vector potential $A_{\mu}$ are ({\it e.g.}, Misner {\it et 
al.} 1973)
\begin{equation}
A^{\mu ;\alpha}_{\verb+  +;\alpha} - R_{\verb+ +\alpha}^{\mu} A^{\alpha} 
= 0 
\end{equation}
when $A^{\mu}$ is chosen to obey the Lorentz gauge condition
\begin{equation}
A^{\mu}_{\verb+ +;\mu} = 0
\end{equation}
It is convenient here to treat $A_{\mu}$ as a complex vector.
Hence the electromagnetic field tensor $F_{\mu \nu}$ is given by 
\begin{equation}
F_{\mu \nu} = {\cal{R}}e(A_{\nu;\mu}- A_{\mu;\nu})
\end{equation}
The corresponding electromagnetic energy-momentum tensor is defined by 
\begin{equation} 
T^{\mu \nu} = \frac{1}{4 \pi} \left[-F^{\mu \rho} F^{\nu}_{.\rho} 
+ \frac{1}{4} F^{\alpha \beta} F_{\alpha \beta} g^{\mu \nu} \right]
\end{equation}
where $F_{. \rho}^{\nu} = g^{\nu \lambda}~F_{\lambda \rho}$. The components of
this tensor satisfy the conservation equations 
$T^{\mu \nu}_{\verb+  +;\nu} = 0$ as a consequence of Eqs.(3). 
\par For an observer located at the spacetime point $x$ and moving with the 
unit 4-velocity $u^{\alpha}$, the density of electromagnetic energy flux is
given by the Poynting vector
\begin{equation}
{\cal{P}}^{\mu}(x,u) = c T^{\mu \nu}(x) u_{\nu}(x)
\end{equation}
and the density of electromagnetic energy as measured by the observer is 
\begin{equation}
\mu_{el}(x,u) =  T^{\mu \nu}(x) u_{\mu} u_{\nu}
\end{equation}
In this paper, we use the geometrical optics approximation. So we assume 
that there exist wave solutions to Eqs.(3) which admit a development of the form
\begin{equation}
A^{\mu}(x,\xi) = [a^{\mu}(x) + O(\xi)] \exp(\frac{i}{\xi} \hat{S}(x))
\end{equation}
where $a^{\mu}(x)$ is a slowly varying complex vector amplitude, $\hat{S}(x)$ 
is a real function and $\xi$ a dimensionless parameter which tends to zero as 
the typical wavelength of the wave becomes shorter and shorter. A solution like
(9) represents a quasi plane, locally monochromatic wave of high frequency  
(Misner {\it et al.} 1973). 
\par Let us define the phase $S$ and the vector field $k_{\alpha}$ by the
relations 
\begin{equation}
S(x,\xi) = \frac{1}{\xi} \hat{S}(x)
\end{equation}
and 
\begin{equation}
k_{\alpha} = S_{, \alpha}
\end{equation} 
Inserting (9) into Eqs.(3) and (4), then retaining only the leading terms of
order $\xi^{-2}$ and $\xi^{-1}$, yield the fundamental 
equations of geometrical optics
\begin{equation}
k^{\alpha} k_{\alpha} = 0
\end{equation}
\begin{equation}
k^{\alpha}a^{\mu}_{; \alpha} = -\frac{1}{2} a^{\mu} k^{\alpha}_{; \alpha}
\end{equation}
with the gauge condition 
\begin{equation}
k_{\alpha} a^{\alpha} = 0
\end{equation}
\par Light rays are defined to be the curves whose tangent vector field is
$k^{\alpha}$. So the parametric equations $x^{\alpha} = x^{\alpha}(v)$ of the
light rays are solutions to the differential equations 
\begin{equation}
\frac{dx^{\alpha}}{dv} = k^{\alpha}(x^{\lambda}(v))
\end{equation}
where $v$ is an affine parameter. Differentiating Eq.(12) and noting that 
\begin{equation}
k_{\alpha;\beta} =k_{\beta ;\alpha}
\end{equation}
follows from (11), it is easily seen that $k^{\alpha}$ satisfies the 
propagation equations 
\begin{equation}
k^{\alpha}k_{\beta;\alpha}=0
\end{equation}
These equations, together with (12), show that the light rays are null
geodesics. 
\par Inserting (9) into (5) and (6) gives the approximate expression for 
$F_{\mu \nu}$  
\begin{equation}
F_{\mu \nu} = {\cal{R}}e[i(k_{\mu} a_{\nu} - k_{\nu} a_{\mu}) e^{iS}] 
\end{equation}
and for $T^{\mu \nu}$ averaged over a period
\begin{equation}
T^{\mu \nu} = \frac{1}{8 \pi} a^2 k^{\mu} k^{\nu}
\end{equation}
where $a$ is the scalar amplitude defined by 
\footnote{We introduce a minus sign in (20) because Eqs.(12) and (14) 
imply that $a^{\mu}$ is a space-like vector when the electromagnetic field is
not a pure gauge field.} 
\begin{equation}
a = (-a^{\mu}\overline{a}_{\mu})^{1/2}
\end{equation}
\par From (7) and (19), it is easily seen that the Poynting vector is 
proportional to the null tangent vector $k^{\mu}$.
This means that the energy of the wave is transported
along each ray with the speed of light. Let us denote by ${\cal{F}}(x,u)$ the
energy flux received by an observer located at $x$ and moving with the
4-velocity $u^{\alpha}$: by definition, ${\cal{F}}(x,u)$ is the amount of
radiating energy flowing per unit proper time across a unit surface orthogonal 
to the direction of propagation. It follows from (8) and (19) that
\begin{equation}
{\cal{F}}(x,u) = c \mu_{el} (x,u) = \frac{c}{8 \pi} a^2(x)
(u^{\mu}k_{\mu})^2_{obs}
\end{equation}
\par This formula enables us to determine the photon flux ${\cal{N}}(x,u)$ 
received by the observer located at $x$ and moving with the 4-velocity 
$u^{\alpha}$. Since the 4-momentum of a
photon is $p^{\mu} = \hbar k^{\mu}$, the energy of the photon as measured by the
observer is $cp^{\mu} u_{\mu} = c\hbar (u^{\mu} k_{\mu})$. We have therefore  
\begin{equation}
{\cal{N}}(x,u)=\frac{1}{8 \pi \hbar} a^2(x) (u^{\mu} k_{\mu})_{obs} 
\end{equation}
The spectral shift $z$ of a light source (emitter) as measured by an observer
is given by ({\it e.g.} G.F.R. Ellis, 1971)
\begin{equation}
1+z = \frac{(u^{\mu} k_{\mu})_{em}}{(u^{\nu} k_{\nu})_{obs}}
\end{equation}
Consequently, the photon flux ${\cal{N}}(x,u)$ may be written as 
\begin{equation}
{\cal{N}}(x,u)=\frac{1}{8 \pi \hbar} a^2(x) 
\frac{( u^{\mu} k_{\mu})_{em}}{1+z} 
\end{equation}
\par The scalar amplitude $a$ can be written in the form of an integral along
the light ray $\gamma$ joining the source to the observer located at $x$. 
Multiplying Eq.(13) by $\overline{a}_{\mu}$ yields the propagation equation 
for $a$
\begin{equation}
k^{\alpha} a_{; \alpha} \equiv \frac{da}{dv} = -\frac{1}{2} a k^{\alpha}_{;
\alpha}
\end{equation}
where $d/dv$ denotes the total differentiation of a scalar function along 
$\gamma$. Then, integrating (25) gives
\begin{equation}
a_{|obs} = a_{|x_0} \exp \left( -\frac{1}{2} \int \limits_{v_{x_0}}^{v_{obs}} 
k^{\alpha}_{; \alpha} \verb+ +dv \right)
\end{equation}
where $x_0$ is an arbitrary point on the light ray $\gamma$. 
\par In the following, we consider that the light source is at spatial
infinity. We suppose the existence of coordinate systems $x^{\alpha}$ such 
that on any 
hypersurface $x^0 = const.$, $|g_{\mu \nu}~-~\eta_{\mu \nu}|~=~O(1/r)$ when 
$r~=~[\sum_{i=1}^3~(x^i)^2]^{1/2}~\rightarrow~\infty$, with 
$\eta_{\mu \nu}~=~diag(1,-1,-1,-1)$. 
We require that in such coordinate systems the quantities $k_{\alpha ; \beta}$, 
$k_{\alpha ; \beta ; \gamma}$ and $a_{;\alpha}$ respectively fulfill the
asymptotic conditions
\begin{equation}
\left\lbrace 
\begin{array}{ccc}
k_{\alpha; \beta} (x_0)          & = &  O( 1/|v_{x_0}|^{1+p}) \\
                                 &   & \\
k_{\alpha; \beta ; \gamma} (x_0) & = &  O(1/|v_{x_0}|^{2+p}) \\
                                 &   & \\
a_{;\alpha} (x_0)                & = &  O(1/|v_{x_0}|^{1+p}) 
\end{array}
\right.
\end{equation}
when $v_{x_0} \rightarrow - \infty$, with $p > 0$. 
Moreover, we assume that the scalar amplitude
$a_{|x_0}$ in Eq.(26) remains bounded when $v_{x_0} \rightarrow - \infty$ and 
we put 
\begin{equation}
\lim_{v_{x_0} \rightarrow - \infty} a_{x_0} = a_0
\end{equation}
It results from these assumptions that $a_{|obs}$ may be written as 
\begin{equation}
a_{|obs} = a_{0} \exp \left( -\frac{1}{2} \int \limits_{-\infty}^{v_{obs}} 
k^{\alpha}_{; \alpha} \verb+ +dv \right)
\end{equation}
\par Now, let us differentiate $k^{\alpha}_{;\alpha}$ with respect to $v$
along $\gamma$. Applying (1) and (2), then taking (16) and (17) into
account, we obtain the relation (Sachs 1961) 
\begin{equation}
\frac{d}{dv}(k^{\alpha}_{;\alpha}) = - k^{\alpha;\beta} k_{\alpha;\beta} -
R_{\alpha \beta} k^{\alpha} k^{\beta} 
\end{equation}
As a consequence, we can write 
\begin{equation}
\int \limits_{-\infty}^{v_{obs}} k^{\alpha}_{; \alpha} \verb+ +dv
 = - \int \limits_{-\infty}^{v_{obs}} dv \int \limits_{-\infty}^{v} 
[R_{\alpha \beta}(x^{\lambda} (v')) k^{\alpha} k^{\beta} +
k^{\alpha;\beta} k_{\alpha;\beta}] dv' 
\end{equation}
The convergence of the integrals is ensured by conditions (27).
\par Equations (29) and (31) allow to determine the factor $a^2(x)$ in
${\cal{N}}(x,u)$ from the energy content of the regions crossed by the light
rays and from the geometry of the rays themselves. 
\par It is well known that $1/(1+z)$ (or $(1+z)$) can also be obtained in the
form of an integral along the light ray $\gamma$ (see {\it e.g.} Ellis 1971 or
Schneider {\it et al.} 1992). However, the corresponding formula will not be
useful for our discussion and we will not develop it here. 
\par In fact, the scintillation phenomenon consists in a variation of
${\cal{N}}$ with respect to time. For this reason, it is more convenient to
calculate the total derivative of ${\cal{N}}$ along the world-line
${\cal{C}}_{obs}$ of a given observer, moving at the point $x$ with the 
4-velocity $u^{\alpha}$. 
\par Given a scalar or tensorial quantity $F$, we denote by $\dot{F}$ the total
covariant differentiation along ${\cal{C}}_{obs}$ defined by
\begin{equation}
\dot{F} \equiv u^{\lambda} F_{;\lambda} = \frac{\nabla F}{ds}
\end{equation} 
where $ds = (g_{\mu \nu} dx^{\mu} dx^{\nu})^{1/2}$ is the line element between
two events $x^{\mu}$ and $x^{\mu}+dx^{\mu}$ on ${\cal{C}}_{obs}$.
\par In Eq.(24), the quantity $c\hbar(u^{\mu}k_{\mu})_{em}$ is the energy of a photon 
emitted by an
atom of the light source as measured by an observer comoving with this atom. So
$(u^{\mu}k_{\mu})_{em}$ is a constant which depends only on the nature of
the atom (this constant characterizes the emitted spectral line). Consequently,
the change in the photon flux with respect to time is simply due to the change
in the scalar amplitude $a$ and to the change in the spectral shift $z$. 
From (24), we obtain at each point $x$ of ${\cal{C}}_{obs}$
\begin{equation}
\frac{\dot{{\cal{N}}}}{{\cal{N}}}  = 2 \frac{\dot{a}}{a} + (1+z) \frac{d}{ds}
\left( \frac{1}{1+z} \right)
\end{equation}
\par Henceforth, we shall call the contribution $2\dot{a}/a$ in Eq.(33) the
geometrical scintillation because the variations in $a$ are related to the
focusing properties of light rays by gravitational fields 
(see G.F.R.Ellis 1971 and references therein; see also Misner {\it et al.} 
1973).
\par Let us now try to find expressions for $\dot{a}/a$ and  $\frac{d}{ds} 
(1+z)^{-1}$ in the
form of integrals along $\gamma$. In what follows, we assume that the ray
$\gamma$ hits at each of its points $x(v)$ a vector field $v^{\mu}$ which
satisfies the boundary condition 
\begin{equation} 
v^{\mu}(x_{obs}) = u^{\mu}_{obs}
\end{equation}
Let us emphasize that $v^{\mu}$ can be chosen arbitrarily at any point $x$
which does not belong to the world line ${\cal{C}}_{obs}$ (for example, 
$v^{\mu}(x)$ could be the unit 4-velocity of an observer at $x$, an
assumption which is currently made in cosmology; however we shall make a more
convenient choice for $v^{\mu}$ in what follows). 
\par It results from the boundary conditions (27) and (34) that $\dot{a}/a$ may
be written as
\begin{equation}
\left.\frac{\dot{a}}{a}\right|_{obs} = \int \limits_{-\infty}^{v_{obs}} \frac{d}{dv} 
[v^{\mu}(\ln a)_{;\mu}] dv
\end{equation}
\par Thus we have to transform the expression 
\begin{equation}
\frac{d}{dv} [v^{\mu}(\ln a)_{;\mu}] = k^{\alpha}( v^{\mu}
(\ln a)_{;\mu})_{;\alpha}
\end{equation}
taken along $\gamma$. Of course, we must take into account the propa\-gation
equation (25) which could be rewritten as 
\begin{equation}
k^{\alpha}(\ln a)_{;\alpha} = - \frac{1}{2} k^{\alpha}_{;\alpha}
\end{equation}
Noting that 
\begin{equation}
k^{\alpha}( v^{\mu} (\ln a)_{;\mu})_{;\alpha} = k^{\alpha} v^{\mu} 
(\ln a)_{;\mu ;\alpha} + k^{\alpha} v^{\mu}_{;\alpha} (\ln a)_{;\mu}
\end{equation}
then using the relation 
\begin{equation}
F_{;\alpha;\beta} =  F_{;\beta;\alpha}
\end{equation}
which holds for any scalar $F$, we find 
\begin{equation}
\frac{d}{dv} [v^{\mu}(\ln a)_{;\mu}] = v^{\mu} [k^{\alpha} (\ln
a)_{;\alpha}]_{;\mu} + [k,v]^{\mu} (\ln a)_{;\mu} 
\end{equation}
where the bracket $[k,v]$ of $k^{\alpha}$ and $v^{\beta}$ is the vector defined 
by  
\begin{equation} 
[k,v]^{\mu} \equiv k^{\alpha} v^{\mu}_{;\alpha} - v^{\alpha} k^{\mu}_{;\alpha}
\end{equation}
Taking (37) into account, it is easily seen that 
\begin{equation} 
\frac{d}{dv} [v^{\mu}(\ln a)_{;\mu}] = -\frac{1}{2} v^{\mu}
k^{\alpha}_{;\alpha ;\mu} + [k,v]^{\mu} (\ln a)_{;\mu} 
\end{equation}
\par Now, using the identity (1) and the definition (2) yields
\begin{equation} 
\frac{d}{dv} [v^{\mu}(\ln a)_{;\mu}] =  \frac{1}{2} R_{\mu \nu} k^{\mu} v^{\nu}
- \frac{1}{2} v^{\mu} k^{\alpha}_{;\mu ;\alpha} +[k,v]^{\mu} (\ln a)_{;\mu} 
\end{equation}
\par Let us try to write the term  
$- \frac{1}{2} v^{\mu} k^{\alpha}_{;\mu ;\alpha}$ in the form of an integral
along $\gamma$. In agreement with (27), we have at any point
$x(v)$ of $\gamma$:
\begin{equation} 
v^{\mu} k^{\alpha}_{;\mu ;\alpha} = \int \limits_{-\infty}^{v}
\frac{d}{dv} (v^{\mu} k^{\alpha}_{;\mu ;\alpha}) dv = \int 
\limits_{-\infty}^{v} k^{\lambda}
(v^{\mu}k^{\alpha}_{;\mu ;\alpha})_{;\lambda}dv 
\end{equation}
\par A tedious but straightforward calculation using (1), (2) and (17) leads
to the following result 
\begin{eqnarray} 
\lefteqn{- \frac{d}{dv}(v^{\mu} k^{\alpha}_{;\mu ;\alpha}) = 
(R_{\rho \sigma ; \mu}-R_{\mu \rho; \sigma}) v^{\mu} k^{\rho} k^{\sigma}} \nonumber \\
& & \verb+           + + R_{\rho \sigma} k^{\rho} v^{\mu}  k^{\sigma}_{;\mu} \\
& & \verb+           + + v^{\mu} (k^{\alpha; \beta} k_{\alpha ; \beta})_{;\mu} 
- [k,v]^{\mu} k^{\alpha}_{;\mu;\alpha} \nonumber
\end{eqnarray}
\par In the above formulae $v^{\mu}$ is an arbitrary vector. So we can
choose $v^{\mu}$ so that the transport equations
\footnote{These equations mean that $v^{\mu} = \alpha \eta^{\mu}$, where
$\alpha=const.$ and $\eta^{\mu}$ is a connection vector of the
system of light rays associated with the phase function $S$ (see, {\it e.g.},
Schneider {\it et al.} 1992).} 
\begin{equation} 
[k,v]^{\mu} = 0
\end{equation}
are satisfied along the ray $\gamma$. Since (46) is a system of first order
partial differential equations in $v^{\mu}$, there exists one and only one
solution satisfying the boundary conditions (34). With this choice, $2
\dot{a}/a$ is given by the integral formula:
\begin{eqnarray} 
\lefteqn{\left. 2\frac{\dot{a}}{a} \right|_{obs} =  \int 
\limits_{-\infty}^{v_{obs}}
R_{\mu \nu} k^{\mu} v^{\nu} dv +  \int \limits_{-\infty}^{v_{obs}} dv  
\int \limits_{-\infty}^{v} 
[(R_{\rho \sigma ; \mu} - R_{\mu \rho ; \sigma})
v^{\mu} k^{\rho} k^{\sigma} } \nonumber \\
& & \verb+                              + +   R_{\rho \sigma} k^{\rho} v^{\mu} k^{\sigma}_{;\mu} \\
\nonumber \\
& & \verb+                              + + v^{\mu} (k^{\alpha ; \beta} k_{\alpha ; \beta})_{;\mu}]dv' 
\nonumber
\end{eqnarray}
\par Now we look for an integral form for the total derivative $\frac{d}{ds}
(1+z)^{-1}$ along ${\cal{C}}_{obs}$. 
Henceforth, we suppose for the sake of simplicity that
the observer is freely falling, {\it i.e.} that ${\cal{C}}_{obs}$ is a timelike
geodesic. So we have
\begin{equation} 
\dot{u}^{\alpha} = u^{\lambda} u^{\alpha}_{;\lambda} = 0
\end{equation}
\par Since $(u^{\mu} k_{\mu})_{em}$ is a constant characterizing the observed
spectral line (see above), it follows from (23) and (48) that
\begin{equation} 
\frac{d}{ds}\left( \frac{1}{1+z} \right)_{obs} = \frac{1}
{(u^{\alpha}k_{\alpha})_{em}} (u^{\mu} u^{\nu} k_{\mu;\nu})_{obs}
\end{equation}
\par Given an arbitrary vector field $v^{\mu}$ fulfilling the boundary
condition (34), Eq.(49) may be written as 
\begin{equation} 
\frac{d}{ds}\left( \frac{1}{1+z} \right)_{obs} = 
\frac{1}{(u^{\alpha}k_{\alpha})_{em}} \int \limits_{-\infty}^{v_{obs}}
k^{\lambda}(v^{\mu} v^{\nu} k_{\mu;\nu})_{;\lambda} dv
\end{equation}
\par Using (1), (17) and (41), a straightforward calculation gives the general
formula
\begin{eqnarray} 
\lefteqn{\frac{d}{ds}\left( \frac{1}{1+z} \right)_{obs} = 
\frac{1}{(u^{\alpha}k_{\alpha})_{em}}
\int \limits_{-\infty}^{v_{obs}} \{-R_{\mu \rho \nu \sigma} v^{\mu} v^{\nu}
k^{\rho} k^{\sigma}} \nonumber \\
 \\
 & & \verb+                + +(k^{\lambda} v^{\mu}_{;\lambda})(v^{\nu} k_{\mu;\nu}) + 
v^{\mu}[k,v]^{\nu} k_{\mu;\nu} \} dv \nonumber
\end{eqnarray}
which holds for any freely falling observer.
\par Now let us choose for $v^{\mu}$ the vector field defined by (46) and (34).
We obtain
\begin{eqnarray} 
\lefteqn{(1+z)\frac{d}{ds} \left( \frac{1}{1+z} \right)_{obs} = 
\frac{1}{(u^{\lambda}k_{\lambda})_{obs}}
\int \limits_{-\infty}^{v_{obs}} [-R_{\mu \rho \nu \sigma} v^{\mu} v^{\nu}
k^{\rho} k^{\sigma}} \nonumber \\
 \\  
 & &\verb+                                + + v^{\mu} v^{\nu} k^{\alpha}_{;\mu} k_{\alpha;\nu}] dv 
\nonumber
\end{eqnarray}
 
\section{Weak-field approximation}
Now we assume the gravitational field to be very weak. So we put 
\begin{equation}
g_{\mu \nu} = \eta_{\mu \nu} + h_{\mu \nu}
\end{equation}
where $h_{\mu \nu}$ is a small perturbation of the flat spacetime
metric $\eta_{\mu \nu}$, and we systematically discard the
terms of order $h^2,h^3,...$ in the following calculations. Thereafter, we
suppose that any quantity $T$ (scalar or tensor) may be written as 
\begin{equation}
T = T^{(0)} + T^{(1)} + O(h^2)
\end{equation}
where $T^{(0)}$ is the unperturbed quantity in flat spacetime and $T^{(1)}$
denotes the perturbation of first-order with respect to $h_{\mu \nu}$. 
Henceforth, indices will be lowered with $\eta_{\mu \nu}$ and
raised with $\eta^{\mu \nu}=\eta_{\mu \nu}$.
\par We shall put for the sake of simplicity 
\begin{equation}
K_{\mu} = k^{(0)}_{\mu} = S^{(0)}_{\verb+ +,\mu} 
\end{equation}
\par Neglecting the first order terms in $h$, Eq.(12) gives 
$K^{\alpha}K_{\alpha} = 0$, whereas Eq.(17) reduces to the equation of a null 
geodesic in flat spacetime related to Cartesian coordinates
\begin{equation}
K^{\alpha} K_{\beta, \alpha} = 0
\end{equation}
\par In agreement with the assumptions made in Sect.3 to obtain Eqs.(29) and 
(31), we
consider that at the zeroth order in $h_{\mu \nu}$, the light emitted by the
source is described by a plane monochromatic wave in a flat spacetime. So we
suppose that the quantities $K_{\mu}$, $a^{(0) \mu}$ and consequently $a^{(0)}$
are constants throughout the domain of propagation. 
\par Moreover, we regard as negligible all the perturbations of gravitational 
origin in the vicinity of the emitter (this hypo\-thesis is natural for a source 
at spatial infinity) and the quantity $a_0$ in Eqs.(28) and (29) is given
consequently by 
\begin{equation} 
a_0 = a^{(0)} = const. 
\end{equation}
Furthermore, it results from $K_{\mu}=const.$ that $k_{\alpha;\beta}~=~O(h)$. 
Therefore, terms like $k^{\alpha}_{;\mu}k_{\alpha;\nu}$ or 
$R_{\rho \sigma}k^{\rho} v^{\mu}  k^{\sigma}_{;\mu}$ are of second order and
can be systematically disregarded. 
\par According to our general assumption in this section, the unit 4-velocity
of the observer may be expanded as 
\begin{equation} 
u^{\alpha}_{obs} = U^{\alpha} + u^{(1)\alpha}_{obs} + O(h^2)
\end{equation}
at any point of ${\cal{C}}_{obs}$, with the definition 
\begin{equation} 
U^{\alpha} = u^{(0)\alpha}_{obs} 
\end{equation}
It follows from (48) and from $g_{\alpha \beta} u^{\alpha} u^{\beta} = 1$ that 
\begin{equation} 
U^{\alpha} = const.
\end{equation}
and 
\begin{equation} 
\eta_{\alpha \beta} U^{\alpha} U^{\beta} = 1
\end{equation}
\par From these last equations, we recover the fact that the unperturbed motion
of a freely falling observer is a time-like straight line in Minkowski
space-time. 
\par Now we have to know the quantities $v^{\mu}$ occurring in Eqs.(47) and (52)
at the lowest order. An elementary calculation shows that, in
Eqs.(46), the covariant differentiation may be replaced by the ordinary
differentiation. So we have to solve the system
\begin{equation} 
\frac{dv^{\alpha}}{dv} = v^{\mu} k^{\alpha}_{,\mu}
\end{equation}
together with the boundary conditions (34). 
\par Assuming the expansion 
\begin{equation} 
v^{\mu} = v^{(0)\mu} + v^{(1)\mu} + O(h^2)
\end{equation}
it is easily seen that the unique solution of (62) and (34) is such that at any
point of the light ray $\gamma$, the components $v^{(0)\mu}$ are constants
given by
\begin{equation} 
v^{(0)\mu} =  U^{\mu}
\end{equation}
\par Neglecting all the second order terms in (47) and (52), we finally obtain
\begin{eqnarray} 
\lefteqn{\left. 2\frac{\dot{a}}{a} \right|_{obs} =  
\int \limits_{-\infty}^{v_{obs}}
 R^{(1)}_{\mu \nu} K^{\mu} U^{\nu} dv} \nonumber \\
 \\
 & & \verb+     + +  \int \limits_{-\infty}^{v_{obs}} dv  \int \limits_{-\infty}^{v} 
(R^{(1)}_{\rho \sigma , \mu} - R^{(1)}_{\mu \rho , \sigma})
U^{\mu} K^{\rho} K^{\sigma} dv' \nonumber 
\end{eqnarray}
and 
\begin{equation} 
(1+z)\frac{d}{ds}\left( \frac{1}{1+z} \right)_{obs} = 
-\frac{1}{U^{\lambda}K_{\lambda}} \int \limits_{-\infty}^{v_{obs}}
R^{(1)}_{\mu \rho \nu \sigma} U^{\mu} U^{\nu} K^{\rho} K^{\sigma} dv
\end{equation}
all the integrations being performed along the unperturbed path of light.
\par In Eq.(66) $R^{(1)}_{\mu \rho \nu \sigma}$ denotes the linearized curvature
tensor of the metric $g_{\mu \nu} = \eta_{\mu \nu} + h_{\mu \nu}$, {\it i.e.} 
\begin{equation} 
R^{(1)}_{\mu \rho \nu \sigma} = - \frac{1}{2}(h_{\mu \nu,\rho \sigma} +h_{\rho
\sigma, \mu \nu} - h_{\mu \sigma , \nu \rho} - h_{\nu \rho , \mu \sigma})
\end{equation}
and $R^{(1)}_{\mu \nu}$ is the corresponding linearized Ricci tensor
\begin{equation}
R^{(1)}_{\mu \nu} = \eta^{\alpha \beta} R^{(1)}_{\alpha \mu \beta \nu}
\end{equation}
\par It is worthy to note that the components $R^{(1)}_{\mu \rho \nu \sigma}$
and $R^{(1)}_{\mu \nu}$ are gauge-invariant quantities. Indeed, under an
arbitrary infinitesimal coordinate transformation $x^{\alpha} \rightarrow
x'^{\alpha} = x^{\alpha}+ \xi^{\alpha}(x)$, $h_{\mu \nu}(x)$ 
transforms into 
$h'_{\mu \nu}(x) = h_{\mu \nu}(x) -\xi_{\mu, \nu} - \xi_{\nu,
\mu}$, and it is easily checked from (67) and (68) that 
\begin{equation}
R^{(1)}_{\mu \rho \nu \sigma}(h'_{\alpha \beta}) =  
R^{(1)}_{\mu \rho \nu \sigma}(h_{\alpha \beta}) 
\end{equation}
\begin{equation}
R^{(1)}_{\mu \nu}(h'_{\alpha \beta}) =
R^{(1)}_{\mu \nu}(h_{\alpha \beta})
\end{equation}
This feature ensures that the right-hand sides of Eqs. (65) and (66) are 
gauge-invariant quantities.
\par Equation (65) reveals that the first order geometrical scintillation
effect depends upon the gravitational field through the Ricci tensor only.
On the other side, it follows from (66) that the part of the scintillation due 
to the spectral shift depends upon the curvature tensor. 
\par These properties have remarkable consequences in general relativity. 
Suppose that the light ray $\gamma$ travels
in regions entirely free of matter. Since the linearized Einstein equations are
in a vacuum
\begin{equation}
R^{(1)}_{\mu \nu} = 0
\end{equation}
it follows from Eq.(65) that 
\begin{equation}
2\frac{\dot{a}}{a} = 0 + O(h^2)
\end{equation}
As a consequence, $\dot{{\cal{N}}}/{\cal{N}}$ reduces to the contribution of 
the change in the spectral shift
\begin{equation}
\frac{\dot{{\cal{N}}}}{{\cal{N}}} = 
-\frac{1}{U^{\lambda}K_{\lambda}} \int \limits_{-\infty}^{v_{obs}}
R^{(1)}_{\mu \rho \nu \sigma} U^{\mu} U^{\nu} K^{\rho} K^{\sigma} dv 
\end{equation}
\par From (72), we recover the conclusion previously drawn by Zipoy (1966) and 
Zipoy \& Bertotti (1968): within general rela\-tivity, gravitational waves
produce no first order geometrical scintillation.

\section{Application to the scalar-tensor theories.}
The general theory developed in the above sections is valid for any metric
theory of gravity. Let us now examine the implications of Eqs.(65) and (66)
within the scalar-tensor theories of gravity. 
\par The class of theories that we consider here is described by the action
\footnote{For details see Will (1993) and
references therein. The factor $-(16 \pi c)^{-1}$ in the gravitational action
is due to the fact that we use the definition of the energy-momentum tensor
given in Landau \& Lifshitz (1975).}
\begin{eqnarray}
\lefteqn{
{\cal{J}} = -\frac{1}{16 \pi c} \int d^4x \sqrt{|g|} \left[\Phi R  -
\frac{\omega(\Phi)}{\Phi} \Phi^{,\alpha} \Phi_{,\alpha} \right ]} \nonumber \\
\\
 & & \verb+  + + {\cal{J}}_{m}(g_{\mu \nu},\psi_m) \nonumber
\end{eqnarray}
where $R$ is the Ricci scalar curvature ($R=g^{\mu \nu}R_{\mu \nu}$), $\Phi$ is
the scalar gravitational field, $g$ is the determinant of the metric components
$g_{\mu \nu}$, $\omega(\Phi)$ is an arbitrary function of the scalar field $\Phi$, and 
${\cal{J}}_{m}$ is the matter action. We assume that ${\cal{J}}_{m}$ is a
functional of the metric and of the matter fields $\psi_m$ only. This means
that ${\cal{J}}_{m}$ does not depend explicitly upon the scalar field $\Phi$
(it is the assumption of universal coupling between matter and metric). 
\par We consider here the weak-field approximation only. So we assume that the
scalar field $\Phi$ is of the form 
\begin{equation}
\Phi = \Phi_0 + \phi 
\end{equation}
where $\phi$ is a first order perturbation of an averaged constant value
$\Phi_0$. Consequently, the field equations deduced from 
(74) reduce to the following system:
\begin{equation}
R^{(1)}_{\mu \nu} = 8 \pi \Phi^{-1}_0 (T^{(0)}_{\mu \nu} -\frac{1}{2}
T^{(0)}\eta_{\mu
\nu}) + \Phi_0^{-1} (\phi_{,\mu \nu} + \frac{1}{2} \Box \phi \eta_{\mu
\nu})
\end{equation}
\begin{equation}
\Box \phi = \frac{8 \pi}{2\omega (\Phi_0) + 3} T^{(0)}
\end{equation}
where $T^{(0)}_{\mu \nu}$ is the energy-momentum tensor of the matter fields 
$\psi_m$ at the lowest order, $T^{(0)}=\eta^{\alpha 
\beta}T^{(0)}_{\alpha \beta}$ and $\Box$ denotes the
d'Alembertian operator on Minkowski spacetime: $\Box \phi = \eta^{\alpha
\beta} \phi_{,\alpha \beta}$.
\par It is easily seen that any solution $h_{\mu \nu}$ to the field equations
(76) is given by \footnote{This transformation can be suggested by the
conformal transformation of the metric which passes from the Jordan-Fierz frame to
the Einstein frame.}
\begin{equation}
h_{\mu \nu} = h^{E}_{\mu \nu} - \frac{\phi}{\Phi_0} \eta_{\mu \nu}
\end{equation} 
where $h^{E}_{\mu \nu}$ is a solution to the equations
\begin{equation}
R^{(1)}_{\mu \nu}(h^{E}_{\alpha \beta}) 
= 8 \pi \Phi^{-1}_0 (T^{(0)}_{\mu \nu} - \frac{1}{2} T^{(0)}\eta_{\mu \nu})
\end{equation} 
which are simply the linearized Einstein equations with an effective
gravitational constant $G_{eff} = c^4 \Phi_0^{-1}$. Indeed, inserting (78) in 
(67) yields the following expression for the curvature tensor 
\begin{eqnarray}
\lefteqn{R^{(1)}_{\mu \rho \nu \sigma}(h_{\alpha \beta}) = 
R^{(1)}_{\mu \rho \nu \sigma}(h^{E}_{\alpha \beta})}  
\nonumber \\
 \\
& &  \verb+       + + \frac{1}{2} \Phi_0^{-1}
(\eta_{\mu \nu} \phi_{,\rho \sigma} + \eta_{\rho \sigma} \phi_{,\mu \nu}
-  \eta_{\mu \sigma} \phi_{,\nu \rho} - \eta_{\nu \rho} \phi_{,\mu \sigma}) 
\nonumber
\end{eqnarray} 
from which one deduces the Ricci tensor 
\begin{equation}
R^{(1)}_{\mu \nu}(h_{\alpha \beta}) = R^{(1)}_{\mu \nu}(h^{E}_{\alpha \beta})
+\Phi_0^{-1} (\phi_{,\mu \nu} + \frac{1}{2} \eta_{\mu \nu} \Box \phi)
\end{equation}
Then substituting for $R^{(1)}_{\mu \nu}(h_{\alpha \beta})$ from its expression
(81) into the field equations (76) gives Eqs.(79), thus proving the proposition.
\par The decomposition (78) of the gravitational perturbation $h_{\mu \nu}$
implies that each term contributing to $\dot{{\cal{N}}}/{\cal{N}}$ can be split
into a part built from the Einsteinian perturbation $h_{\mu \nu}^{E}$ only
and into an other part built from the scalar field $\phi$ alone. In what
follows, we use the superscript ${ST}$ for a functional of a solution $(h_{\mu
\nu},\phi)$ to the field equations (76)-(77) and the superscript ${E}$ for the same
kind of functional evaluated only with the corresponding solution $h_{\mu
\nu}^{E}$.
\par In order to perform the calculation of the integrals (65) and (66), we note
that $K^{\mu} F_{,\mu}$ is the usual total derivative of the quantity $F$ along
the unperturbed ray path, which implies that 
\begin{equation}
\int \limits_{-\infty}^{v_{obs}}  K^{\mu} F_{,\mu} dv = F_{obs} - F_{(-\infty)}
\end{equation}
\par The 4-vector $K^{\mu}$ (supposed here to be future oriented, {\it i.e}
such that $K^0 > 0$) gives the direction of propagation of the light coming from
the observed source. For a given observer moving with the 4-velocity $U^{\mu}$, 
let us put 
\begin{equation}
 N^{\mu} = (\eta^{\mu \nu} - U^{\mu} U^{\nu})
\frac{K_{\nu}}{(U^{\lambda}K_{\lambda})} = \frac{K^{\mu}}
{(U^{\lambda}K_{\lambda})}- U^{\mu}
\end{equation}
We have $\eta^{\mu \nu} N_{\mu}N_{\nu} = -1$. Since $N^{\mu}$ is orthogonal to
$U^{\mu}$ by construction, $N^{\mu}$ can be identified to the unit 3-vector
$\vec{N}$ giving the direction of propagation of the light ray in the usual
3-space of the observer. 
\par Using (76), (77), (80) and the assumption $\phi_{(-\infty)} = 0$, it is 
easily seen that (65) and (66) can be respectively written as 
\begin{equation}
\left. 2\frac{\dot{a}}{a} \right|^{ST}_{obs} = 
\left. 2\frac{\dot{a}}{a} \right|^{E}_{obs} + \left.
\frac{\dot{\phi}}{\Phi_0}\right|_{obs}
\end{equation}
and 
\begin{eqnarray}
\lefteqn{(1+z)\frac{d}{ds}\left( \frac{1}{1+z} \right)^{ST}_{obs} =
(1+z)\frac{d}{ds}\left( \frac{1}{1+z} \right)^{E}_{obs}} \nonumber \\
 \\
 & &\verb+                  ++ \frac{1}{2\Phi_0} (\dot{\phi} - \vec{N}.\vec{\nabla} \phi)_{obs} \nonumber
\end{eqnarray}
where
\begin{equation}  
\dot{\phi} = U^{\mu} \phi_{,\mu}
\end{equation} 
and 
\begin{equation}  
\vec{N}.\vec{\nabla} \phi=N^{\mu}\phi_{,\mu}
\end{equation}  
\par As a consequence, the rate of variation in the photon flux as received by 
the observer is given by the general formula 
\begin{equation}
\left.\frac{\dot{{\cal{N}}}}{{\cal{N}}}\right|^{ST}_{obs} = 
\left.\frac{\dot{{\cal{N}}}}{{\cal{N}}}\right|^{E}_{obs}  +
\frac{1}{2 \Phi_0}(3 \dot{\phi} - \vec{N}.\vec{\nabla} \phi)_{obs}
\end{equation}
\par In a vacuum $(T^{(0)}_{\mu \nu} = 0)$, the metric $h^{E}_{\mu \nu}$
satisfies the linearized Einstein field equations (71) and Eq.(84) reduces to
\begin{equation}
2\left.\frac{\dot{a}}{a} \right|^{ST}_{obs} = 
\left. \frac{\dot{\phi}}{\Phi_0}\right|_{obs}
\end{equation}
\par In Eq.(88), $({\dot{\cal{N}}}/{\cal{N}})^E_{obs}$ is reduced to the term
given by Eq.(73), where $R^{(1)}_{\mu \rho \nu \sigma}$ is constructed with
$h^{E}_{\mu \nu}$.
\par It follows from (89) that contrary to general relativity, the scalar-tensor
theories (defined by (74)) predict the existence of a first-order geometrical
scintillation effect produced by gravitational waves. This effect is
proportional to the amplitude of the scalar perturbation. It should be noted
that an effect of the same order of magnitude is also due to the change in the 
spectral shift.
\par To finish, let us briefly examine the case where the scalar wave $\phi$ is
locally plane (it is a reasonable assumption if the source of gravitational
wave is far from the observer). Thus we can put in the vicinity of the observer
located at the point $x_{obs}$
\begin{equation} 
\phi = \phi(u)
\end{equation}
where $u$ is a phase function which admits the expansion
\begin{equation} 
u(x) = u(x_{obs}) + L_{\mu}(x^{\mu}-x^{\mu}_{obs}) 
+ O(|x^{\mu}-x^{\mu}_{obs}|^2)
\end{equation}
with
\begin{equation}
L_{\mu} = const.
\end{equation} 
\par It follows from Eq.(77) with $T^{(0)} = 0$ that $L_{\mu}$ is a null vector
of Minkowski spacetime. 
\par Replacing $K_{\mu}$ by $L_{\mu}$ in (83) defines the
spacelike vector $P^{\mu}$, which can be identified with the unit 3-vector
$\vec{P}$ giving the direction of propagation of the scalar wave in the
3-space of the observer. Then introducing the angle $\theta$ between
$\vec{N}$ and $\vec{P}$, a simple calculation yields
\begin{equation}
\left.\frac{\dot{{\cal{N}}}}{{\cal{N}}}\right|^{ST}_{obs} = 
\left.\frac{\dot{{\cal{N}}}}{{\cal{N}}}\right|^{E}_{obs} +
(1 + \cos^2\frac{\theta}{2}) \left. \frac{\dot{\phi}}{\Phi_0}
\right|_{obs}
\end{equation} 
\par This formula shows that the contribution of the scalar wave to the
scintillation cannot be zero, whatever be the direction of observation of the 
distant light source.

\section{Are observational tests possible?}
\par It follows from our formulae that the scintillation effects specifically
predicted by scalar-tensor theories are proportional to the amplitude of the
scalar field perturbation at the observer. 
This {\it local} character casts a serious doubt on the detectability of these
effects, since the scalar field perturbation is very small for a localized 
source of gravitational waves.  
\par Indeed, one can put in most cases $\phi/\Phi_0 \sim \alpha^2 h$, where
$\alpha^2$ is a dimensionless constant coupling the scalar field with the
metric gravitational field (see, {\it e.g.}, Damour \& Esposito-Far\`ese 1992).
Experiments in the solar system and observations of binary pulsars like 
PSR 1913+16 indicate that $\alpha^2<10^{-3}$. 
Consequently, setting 
$h \sim 10^{-22}$ for gravitational waves emitted by localized sources gives
$\phi/\Phi_0 < 10^{-25}$ and the effect is much too weak to be detected. 

The authors would like to acknowledge V.Faraoni for useful
discussions. A helpful comment of the referee on the orders of magnitude is
also acknowledged. Finally, one of the authors (C.B.) would like to thank 
G.Esposito-Far\`ese for stimulating discussions.


\end{document}